\documentstyle[aps,eqsecnum,floats]{revtex}
\begin{document}
\draft
\twocolumn[\hsize\textwidth\columnwidth\hsize\csname 
           @twocolumnfalse\endcsname
\title{Gravitational waves from inspiraling compact binaries: \\
       Validity of the stationary-phase approximation to the Fourier transform}
\author{Serge Droz$^*$, Daniel J.~Knapp$^\dag$, Eric Poisson}
\address{Department of Physics, University of Guelph, Guelph,
         Ontario, Canada N1G 2W1}
\author{Benjamin J.~Owen}
\address{Theoretical Astrophysics 130-33, California Institute of 
         Technology, Pasadena, California 91125, and\\
         Max Planck Institut f\"ur Gravitationsphysik, Schlaatzweg 1,
         D-14473 Potsdam, Germany}
\date{Draft of January 26, 1999}
\maketitle
\begin{abstract}
We prove that the oft-used stationary-phase method gives a very
accurate expression for the Fourier transform of the
gravitational-wave signal produced by an inspiraling compact
binary. We give three arguments. First, we analytically calculate the
next-order correction to the stationary-phase approximation, and show
that it is small. This calculation is essentially an application of
the steepest-descent method to evaluate integrals. Second, we 
numerically compare the stationary-phase expression to the results 
obtained by Fast Fourier Transform. We show that the differences can 
be fully attributed to the windowing of the time series, and that they 
have nothing to do with an intrinsic failure of the stationary-phase
method. And third, we show that these differences are negligible for
the practical application of matched filtering. 
\end{abstract}
\vskip 2pc]

\narrowtext

\section{Introduction and summary} 

The ongoing construction of kilometer-size interferometric
gravitational-wave detectors, such as the American LIGO (Laser
Interferometer Gravitational-wave Observatory), the French-Italian
VIRGO, and the British-German GEO600, has motivated a lot of recent
work on strategies to analyze the data to search for and measure the
gravitational-wave signals \cite{1,2,3,4,5,6,7,8,9,10,11,12,13,14,15,OS}.
Much of this work has been devoted to inspiraling compact binaries,
composed of neutron stars and/or black holes, which are one of the
most promising sources of gravitational waves for these detectors. The
key idea behind this work is that the gravitational-wave signals will
be detected and measured by matched filtering \cite{16}, a well-known
technique by which the detector's data stream is cross-correlated with
a set of model waveforms (called templates), and the signal-to-noise
ratio maximized as a function of the template parameters.

While the detector output is represented by a discrete time series,
the operations associated with matched filtering are usually carried
out in (discrete) frequency space, which requires taking the
(discrete) Fourier transform of the time series; the method of choice
here is the standard Fast Fourier Transform (FFT). At the same time,
it is necessary to compute the Fourier transform of $h(t)$, the model
waveform. The waveform $h(t)$ can easily be discretized to mimic the
discrete sampling of the detector output, and the discrete time series
can easily be Fourier transformed by FFT to yield the discrete
analogue of the (continuous) Fourier transform $\tilde{h}(f)$.

However, in theoretical 
investigations \cite{1,2,3,4,5,6,7,8,9,10,11,12,13,15,OS} 
it is often much more convenient to deal with $\tilde{h}(f)$ as an
analytic expression rather than as a table of values; efforts to
obtain such an analytical expression, even if it is only approximate,
are therefore well justified. It can also be argued that to express
the waveform in the frequency domain is in a sense more natural than
to express it in the time domain. One of the reasons is that the
orbital energy, whose expression is required in the derivation of the
waveform, is primarily a function of $F$, the gravitational wave's
instantaneous frequency (which will defined precisely below); it is
therefore natural to express $h$ also as a function of $F$, which
unlike $t$ is a coordinate-independent quantity. Another reason
resides with the fact that the post-Newtonian expression for the
relation $F(t)$ suggests that the instantaneous frequency is not
always a monotonically increasing function of time \cite{GRASP}, a
nonphysical behavior that signals the eventual breakdown of the
post-Newtonian expansion; in contrast, the post-Newtonian expression
for the inverted relation $t(F)$ {\it is} monotonic, a property that 
suggests that $F$ is indeed a more natural time variable. 

In the past, and in the specific context of inspiraling binaries,
analytic expressions for $\tilde{h}(f)$ have been obtained within the
stationary-phase approximation \cite{17}, essentially the
leading-order term in an expansion in powers of the small quantity
(radiation-reaction time scale)/(orbital period). While it has
generally been felt that this approximation is adequate, this belief
has not yet been backed up (in the published literature) with a detailed
quantitative analysis. Claims in the literature \cite{14} to the
effect that this approximation is {\it not} adequate have prompted us
to examine this question.

Our objective in this paper is to prove, beyond any reasonable doubt,
that the stationary-phase approximation to the Fourier transform of an
inspiraling-binary signal is in fact very accurate. First, we
calculate the next-order correction to the stationary-phase
approximation, and show that it is indeed small. This calculation is
essentially an application of the steepest-descent method to evaluate
integrals \cite{17}. We then numerically compare the stationary-phase 
expression for $\tilde{h}(f)$ to the results obtained by FFT. We show 
that the differences can be fully attributed to the windowing of the time
series, and that they are irrelevant for matched filtering. 

We now present and explain our main results. As we shall justify
below, it is sufficient for our purposes to consider the leading-order
expression for the gravitational-wave signal, obtained by assuming
that the binary's orbital motion (taken to be circular) is governed by
Newtonian gravity, with an inspiral --- caused by the loss of energy
and angular momentum to gravitational radiation --- governed by the
Einstein quadrupole formula. This so-called {\it Newtonian} signal is
given by \cite{1}
\begin{equation}
h(t) = Q(\mbox{angles})\, \frac{{\cal M}}{r}\, (\pi {\cal M} F)^{2/3}\,
e^{-i \Phi},
\label{1.1}
\end{equation}
where $Q$ is a function of all the relevant angles (position of the
source in the sky, inclination of the orbital plane, orientation of
the gravitational-wave detector), ${\cal M} = 
(m_1 m_2)^{3/5}/(m_1+m_2)^{1/5}$ (with $m_1$ and $m_2$ denoting the
individual masses) is the {\it chirp mass}, $r$ is the distance to the
source, $F(t)$ the instantaneous frequency (twice the orbital
frequency), and $\Phi(t) = \int 2\pi F(t')\, dt'$ is the phase. The
function $h(t)$ represents the gravitational wave measured at the
detector site. It is a linear superposition of the wave's two
fundamental polarizations, and we choose to express it in this complex
form for convenience. The relation between $F$ and $t$ is
given by \cite{1}
\begin{equation}
\pi {\cal M} F(t) = \biggl[ \frac{5 {\cal M}}{256(t_c-t)}
                    \biggr]^{3/8}, 
\label{1.2}
\end{equation}
where $t_c$, the ``time at coalescence'', is such that formally,
$F(t_c) = \infty$. The relation between $\Phi$ and $t$ is
\begin{equation}
\Phi(t) = \Phi_c - \frac{1}{16}\, 
\biggl[ \frac{256(t_c-t)}{5 {\cal M}} \biggr]^{5/8}, 
\label{1.3}
\end{equation}
where $\Phi_c$, the ``phase at coalescence'', is equal to 
$\Phi(t_c)$. The Newtonian signal is the leading-order term in the
expansion of the gravitational waves in powers of $V \ll 1$, where 
$V = (\pi {\cal M} F)^{1/3}$ is (up to a numerical factor) the orbital
velocity. Post-Newtonian corrections to this result come with a relative
factor of order $V$ in the amplitude, and a relative factor of order
$V^2$ in the phase \cite{18}. Throughout this paper we use geometrized 
units, setting $G=c=1$.

In the stationary-phase approximation, the Fourier transform of the 
function $h(t)$ appearing in Eq.~(\ref{1.1}) is given by \cite{5}
\begin{equation}
\tilde{h}_{\rm spa}(f) = \frac{\sqrt{30\pi}}{24}\, \frac{Q {\cal M}^2}{r}\,
v^{-7/2} e^{i \psi},
\label{1.4}
\end{equation}
where $v \equiv (\pi {\cal M} f)^{1/3} \ll 1$, and 
\begin{equation}
\psi(v) = 2\pi f t_c - \Phi_c - \frac{\pi}{4} + \frac{3}{128v^5}.
\label{1.5}
\end{equation}
Two types of corrections to this result are calculated in Sec.~II. 

The first type of correction constitutes an intrinsic improvement on
the stationary-phase approximation. We show that the steepest-descent 
evaluation of the Fourier transform returns the same amplitude as
before, but that the phase is altered by a term $\delta \psi$:
$\psi \to \psi + \delta \psi$, where
\begin{equation}
\delta \psi(v) = \frac{92}{45}\, v^5 + O(v^{10}).
\label{1.6}
\end{equation}
Notice that this is a correction of order $v^{10}$ {\it relative} to
$\psi(v)$. This is much smaller than post-Newtonian corrections to the
phase, which appear at relative order $v^2$ \cite{5}. Incorporating
these post-Newtonian corrections into our calculation would only
change Eq.~(\ref{1.6}) by adding a term of order $v^7$ to the
right-hand side.  This justifies the fact that it was not necessary, 
for the purposes of this investigation, to use more accurate versions
of Eqs.~(\ref{1.2}) and (\ref{1.3}).

The second type of correction addresses an implicit assumption of the
stationary-phase method, that the function $h(t)$ has support in the
complete time interval $-\infty < t < t_c$. Physically, this
assumption means that the binary system must have formed in the
infinite past (a reasonable assumption given the long lives of compact
binaries), and that the inspiral must continue until $F = F(t_c) =
\infty$ (an unrealistic assumption).  If we choose instead to restrict
the time interval to $t_{\rm min} < t < t_{\rm max}$, such that
$F_{\rm min} \equiv F(t_{\rm min}) > 0$ and $F_{\rm max} \equiv
F(t_{\rm max}) < \infty$, then the Fourier transform will be affected,
and it will differ from $\tilde{h}_{\rm spa}(f)$. The value of $F_{\rm
  min}$ is typically chosen to reflect the lower bound of the
instrument's frequency band.  The value of $F_{\rm max}$ could be
chosen to reflect the upper bound of the instrument's frequency band,
or for some binaries it can be chosen to correspond to the
approximate frequency of the last stable orbit, at which the inspiral
signal changes over to a poorly-known merger signal.  

We shall refer
to this truncation of the time interval as ``windowing'', and for
concreteness, we will assume that the signal is started abruptly at $t =
t_{\rm min}$ and ended abruptly at $t = t_{\rm max}$. Thus, $h(t)$ is
assumed to be given by Eq.~(\ref{1.1}) in the interval $t_{\rm min} <
t < t_{\rm max}$, and is assumed to be zero outside this interval. In
Sec.~II we show that windowing affects both the amplitude and the
phase of the Fourier transform. The amplitude acquires an extra factor
$1 + \delta A_{\rm w}$, where
\begin{eqnarray}
\delta A_{\rm w}(v) &=& -\frac{12}{\sqrt{30\pi}}\, v^{7/2} \biggl[ 
\frac{{x_{\rm min}}^2}{v^3 - {x_{\rm min}}^3}\,  
\cos(\phi_{\rm min} + {\textstyle \frac{\pi}{4}}) 
\nonumber \\ & & \mbox{}
+ \frac{{x_{\rm max}}^2}{{x_{\rm max}}^3 - v^3}\,
\cos(\phi_{\rm max} + {\textstyle \frac{\pi}{4}}) \biggr],
\label{1.7}
\end{eqnarray}
where $x_{\rm min} \equiv (\pi {\cal M} F_{\rm min})^{1/3}$ and 
\begin{equation}
\phi_{\rm min} = \frac{5 v^3 - 8 {x_{\rm min}}^3}{128 {x_{\rm min}}^8}
+ \frac{3}{128 v^5},
\label{1.8}
\end{equation}
with similar equations holding for $x_{\rm max}$ and $\phi_{\rm max}$. 
On the other hand, the phase acquires an extra term 
$\delta \psi_{\rm w}$ given by
\begin{eqnarray}
\delta \psi_{\rm w}(v) &=& \frac{12}{\sqrt{30\pi}}\, v^{7/2} \biggl[ 
\frac{{x_{\rm min}}^2}{v^3 - {x_{\rm min}}^3}\,  
\sin(\phi_{\rm min} + {\textstyle \frac{\pi}{4}}) 
\nonumber \\ & & \mbox{}
+ \frac{{x_{\rm max}}^2}{{x_{\rm max}}^3 - v^3}\,
\sin(\phi_{\rm max} + {\textstyle \frac{\pi}{4}}) \biggr].
\label{1.9}
\end{eqnarray}
The fact that $\delta A_{\rm w}$ and $\delta \psi_{\rm w}$ both
diverge at the boundaries $v = x_{\rm min}$ and $v = x_{\rm max}$
signals the breakdown of our approximations there. Away from the
boundaries, $\delta A_{\rm w}$ and $\delta \psi_{\rm w}$ are bounded,
and they oscillate as a function of frequency. Thus, these corrections
represent amplitude and phase modulations induced by the abrupt
cutoffs of the function $h(t)$ at the boundary points. This is to be
contrasted with $\delta \psi$, which represents a steady phase drift. 

Our complete expression for the steepest-descent approximation to the
Fourier transform is therefore
\begin{equation}
\tilde{h}_{\rm sda}(f) = \tilde{h}_{\rm spa}(f) (1 + \delta A_{\rm w})
e^{i(\delta \psi + \delta \psi_{\rm w})}.
\label{1.10}
\end{equation}
In Sec.~III we show that the only noticeable corrections to 
$\tilde{h}_{\rm spa}(f)$ are the amplitude and phase modulations that
come as a consequence of windowing; in particular, the intrinsic
correction $\delta \psi$ is too small to be noticeable in the relevant
frequency interval. We do this by comparing $\tilde{h}_{\rm spa}(f)$
to $\tilde{h}_{\rm fft}(f)$, the discrete Fourier transform of the
windowed time series $h(t)$; this comparison reveals that any
discrepancy between the two versions of the Fourier transform can be
fully accounted for by the modulations $\delta A_{\rm w}$ and 
$\delta \psi_{\rm w}$. This allows us to conclude that windowing, and
windowing only, must be held responsible for any discrepancy between 
$\tilde{h}_{\rm spa}(f)$ and $\tilde{h}_{\rm fft}(f)$. While limited
to the Newtonian signal of Eq.~(\ref{1.1}), there is no reason to
believe that this conclusion would be invalidated by a full
post-Newtonian analysis \cite{Gopa}.  

Finally, in Sec.~IV we calculate the matched-filtering overlap
between $\tilde{h}_{\rm spa}(f)$ and $\tilde{h}_{\rm fft}(f)$, and
show that the modulations do not significantly affect the
overlap. This result, together with our previous findings, lead us to
conclude that for the purposes of matched filtering, 
$\tilde{h}_{\rm spa}(f)$ and $\tilde{h}_{\rm fft}(f)$ are essentially
{\it equivalent} representations of the gravitational-wave signal. 

The main conclusion of this work is that the stationary-phase method
returns a sufficiently accurate expression for $\tilde{h}(f)$. What's
more, from the fact that $\delta\psi/\psi = O(v^{10})$, we can be sure
that the method will stay accurate for as long as the post-Newtonian
expansion of $h(t)$ in powers of $V$ is itself an accurate
approximation to the gravitational-wave signal. 

\section{Calculation of the Fourier transform}

We begin with the time-domain signal of Eq.~(\ref{1.1}), and we assume
that the signal begins abruptly at a time $t_{\rm min}$ and ends
abruptly at a time $t_{\rm max}$. We let $F_{\rm min} \equiv 
F(t_{\rm min})$ and $F_{\rm max} \equiv F(t_{\rm max})$ be the
corresponding instantaneous frequencies. The relation between $F$ and
$t$ is obtained by integrating
\begin{equation}
\frac{dF}{dt} = \frac{96}{5\pi {\cal M}^2}\, (\pi {\cal M} F)^{11/3},
\label{2.2}
\end{equation}
which leads to
\begin{equation}
t(F) = t_c - \frac{5 {\cal M}}{256}\, (\pi {\cal M} F)^{-8/3},
\label{2.3}
\end{equation}
where $t_c$ (``time at coalescence'') is a constant of
integration. The phase function is then given by
\begin{equation}
\Phi(F) = \Phi_c - \frac{1}{16}\, (\pi {\cal M} F)^{-5/3},
\label{2.4}
\end{equation}
where $\Phi_c$ (``phase at coalescence'') is another constant of 
integration. 

The Fourier transform,
\begin{equation}
\tilde{h}(f) = \int h(t) e^{2\pi i f t}\, dt,
\label{2.5}
\end{equation}
is evaluated by introducing a new integration variable,
\begin{equation}
x \equiv (\pi {\cal M} F)^{1/3},
\label{2.6}
\end{equation}
which can be related to $t$ via Eqs.~(\ref{2.2}) and (\ref{2.3}).
After some re-arrangement, we obtain
\begin{eqnarray}
\tilde{h}(f) &=& \frac{5 Q {\cal M}^2}{32 r}\,
\exp\biggl[ i \biggl( 2\pi f t_c - \Phi_c + \frac{3}{128 v^5} \biggr) 
\biggr] \nonumber \\
& & \mbox{} \times 
I(v,x_{\rm min},x_{\rm max}),
\label{2.7}
\end{eqnarray}
where $x_{\rm min} = (\pi {\cal M} F_{\rm min})^{1/3}$,
$x_{\rm max} = (\pi {\cal M} F_{\rm max})^{1/3}$, and  
\begin{equation}
v \equiv (\pi {\cal M} f)^{1/3} \ll 1.
\label{2.8}
\end{equation}
We have introduced the Fourier integral
\begin{equation}
I(v,x_{\rm min},x_{\rm max}) = \int_{x_{\rm min}}^{x_{\rm max}}
x^{-7} e^{-i \phi}\, dx,
\label{2.9}
\end{equation}
where
\begin{equation}
\phi(x;v) = \frac{5 v^3 - 8x^3}{128 x^8} + \frac{3}{128 v^5}.
\label{2.10}
\end{equation}
It is easy to check that $\phi(v;v) = \phi'(v;v) = 0$, where a prime
denotes differentiation with respect to $x$. The function $\phi(x;v)$
initially decreases from $\phi_{\rm min}(v) \equiv \phi(x_{\rm
min};v)$ to zero as $x$ increases from $x_{\rm min}$ to $v$, and then
increases from zero to $\phi_{\rm max}(v) \equiv \phi(x_{\rm max};v)$
as $x$ increases from $v$ to $x_{\rm max}$. 

The Fourier integral is evaluated by using $\phi$ as the integration 
variable. Because $\phi$ is not monotonic in the interval $[x_{\rm min},
x_{\rm max}]$, the integral must be broken down into two parts. The first 
part covers the interval $[x_{\rm min},v)$, while the second part covers
the interval $(v,x_{\rm max}]$. It is easy to check that the integral
can be expressed as 
\begin{equation}
I(v,x_{\rm min},x_{\rm max}) = \frac{16}{5}\, \Bigl[ J_1(v,x_{\rm min}) 
+ J_2(v,x_{\rm max}) \Bigr],
\label{2.11}
\end{equation}
where
\begin{equation}
J_1(v,x_{\rm min}) = \int_0^{\phi_{\rm min}} \frac{x^2}{v^3-x^3}\,
e^{-i\phi}\, d\phi
\label{2.12}
\end{equation}
and
\begin{equation}
J_2(v,x_{\rm max}) = \int_0^{\phi_{\rm max}} \frac{x^2}{x^3-v^3}\,
e^{-i\phi}\, d\phi.
\label{2.13}
\end{equation}
We recall that $\phi_{\rm min} = \phi(x_{\rm min};v)$ and 
$\phi_{\rm max} = \phi(x_{\rm max};v)$ are functions of $v$. 

The coordinate transformation $x \to \phi$ is defined implicitly
by Eq.~(\ref{2.10}). It will prove sufficient to invert this relation 
in a neighborhood of $x=v$, or $\phi = 0$. The following relations are 
established by Taylor expansion:
\begin{eqnarray}
x &=& v 
\mp \frac{4\sqrt{30}}{15}\, v^{7/2} \phi^{1/2} 
+ \frac{256}{45}\, v^6 \phi 
\mp \frac{10016\sqrt{30}}{2025}\, v^{17/2} \phi^{3/2}
\nonumber \\ & & \mbox{}
+ \frac{4281344}{30375}\, v^{11} \phi^2
+ O(v^{27/2} \phi^{3/2}),
\label{2.14}
\end{eqnarray}
where the upper sign refers to the interval $[x_{\rm min},v)$, while the 
lower sign refers to $(v,x_{\rm max}]$. We also have
\begin{eqnarray}
\frac{x^2}{v^3 - x^3} &=& \frac{\sqrt{30}}{24}\, v^{-7/2} \phi^{-1/2}
+ \frac{5}{9}\, v^{-1} 
- \frac{23\sqrt{30}}{135}\, v^{3/2} \phi^{1/2}
\nonumber \\ & & \mbox{}
+ \frac{19232}{6075}\, v^4 \phi 
+ O(v^{13/2} \phi^{3/2})
\label{2.15}
\end{eqnarray}
in the first interval, and 
\begin{eqnarray}
\frac{x^2}{x^3 - v^3} &=& \frac{\sqrt{30}}{24}\, v^{-7/2} \phi^{-1/2}
- \frac{5}{9}\, v^{-1} 
- \frac{23\sqrt{30}}{135}\, v^{3/2} \phi^{1/2}
\nonumber \\ & & \mbox{}
- \frac{19232}{6075}\, v^4 \phi 
+ O(v^{13/2} \phi^{3/2})
\label{2.16}
\end{eqnarray}
in the second interval. 

To evaluate the integrals $J_1(v,x_{\rm min})$ and $J_2(v,x_{\rm max})$, 
we let $\phi = \alpha - i \beta$ and deform the contour of integration 
into the complex plane. While the original contour is along the
$\alpha$ axis, from 0 to $\phi_{\rm min}$ or $\phi_{\rm max}$,
we take the new contour to be the union of $C$ and $C'$, where $C$ is
the curve $\alpha = 0$ with $\beta$ running from $0$ to $\infty$, while
$C'$ is the curve $\alpha = \phi_{\rm min}\ \mbox{or}\ \phi_{\rm max}$
with $\beta$ running from $\infty$ back to $0$. The contour is completed 
by joining $C$ and $C'$ with the curve $\beta = \infty$, with $\alpha$ 
running from $0$ to $\phi_{\rm min}$ or $\phi_{\rm max}$; this part of 
the contour does not contribute to the integral. The advantage of this
choice of contour is that the integrand is exponentially suppressed away 
from $\beta = 0$, ensuring a rapid convergence of the integral. 

We evaluate the contribution from $C$ to $J_1(v,x_{\rm min})$ by
substituting Eq.~(\ref{2.15}) into (\ref{2.12}), replacing $\phi$
by $-i\beta$ and using $\beta=0$ and $\beta = \infty$ as limits.
The integrations give rise to $\Gamma$-functions, and we obtain
\begin{eqnarray}
J_1^{C} &=& \frac{\sqrt{-30i\pi}}{24}\, v^{-7/2}
- \frac{5i}{9}\, v^{-1} 
+ \frac{23\sqrt{30i\pi}}{270}\, v^{3/2}
\nonumber \\ & & \mbox{}
- \frac{19232}{6075}\, v^4 
+ O(v^{13/2}).
\label{2.17}
\end{eqnarray}
A similar calculation also reveals 
\begin{eqnarray}
J_2^{C} &=& \frac{\sqrt{-30i\pi}}{24}\, v^{-7/2}
+ \frac{5i}{9}\, v^{-1} 
+ \frac{23\sqrt{30i\pi}}{270}\, v^{3/2}
\nonumber \\ & & \mbox{}
+ \frac{19232}{6075}\, v^4 
+ O(v^{13/2}).
\label{2.18}
\end{eqnarray}

The contribution from $C'$ to $J_1(v,x_{\rm min})$ is calculated 
by letting $\phi = \phi_{\rm min} - i\beta$, and expressing the 
function $f(\phi) \equiv x^2/(v^3-x^3)$ as a Taylor series about 
$\phi_{\rm min}$. Thus, $f(\phi_{\rm min} - i\beta) = 
f(\phi_{\rm min}) - i f'(\phi_{\rm min}) \beta + \cdots$ is
substituted into Eq.~(\ref{2.12}), whose limits are replaced by $\beta
= \infty$ and $\beta = 0$. The resulting integrations are again
elementary, and we obtain 
\begin{equation}
J_1^{C'} = i e^{-i \phi_{\rm min}} g(v,x_{\rm min}),
\label{2.19}
\end{equation}
where
\begin{equation}
g(v,x) = \frac{x^2}{\bigl| v^3 - x^3 \bigr|} + \frac{16i}{5}\,
\frac{x^{10} (x^3 + 2v^3)}{\bigl| v^3 - x^3 \bigr|^3} + \cdots.
\label{2.20}
\end{equation}
A similar calculation also reveals 
\begin{equation}
J_2^{C'} = i e^{-i \phi_{\rm max}} g(v,x_{\rm max}).
\label{2.21}
\end{equation}
We note that when $x \ll 1$, the function $g(v,x)$ can be well
approximated by its first term, $x^2/|v^3 - x^3|$, except when $x
\simeq v$. In this situation, the expansion for $g(v,x)$ does not
converge, and our method of calculation breaks down. Thus, our
expression for the Fourier transform will be accurate only when $v^3
\equiv \pi {\cal M} f$ is not too close to either ${x_{\rm min}}^3
\equiv \pi {\cal M} F_{\rm min}$ or ${x_{\rm max}}^3 \equiv 
\pi {\cal M} F_{\rm max}$. In other words, our expression will be
inaccurate near the boundaries $f=F_{\rm min}$ and $f=F_{\rm max}$. 

It is easy to see from Eqs.~(\ref{2.19}) and (\ref{2.20}) that the 
contribution from $C'$ to $J_1(v,x_{\rm min})$ vanishes in the limit 
$x_{\rm min} \to 0$. Similarly, it can be shown that the contribution 
from $C'$ to $J_2(v,x_{\rm max})$ vanishes (as ${x_{\rm max}}^{-6}$)
in the limit $x_{\rm max} \to \infty$. [This behavior is not revealed
by Eq.~(\ref{2.20}), which gives $g(v,x)$ as a series expansion for
small values of $x$. An alternative expression for $g(v,x)$,
appropriate for large values of $x$, can easily be obtained.] This
allows us to conclude that if the boundaries are pushed to 
$F_{\rm min} = 0$ and $F_{\rm max} = \infty$, then $C'$ no longer
contributes to $J_1(v,x_{\rm min})$ and $J_2(v,x_{\rm max})$.

Gathering the results, Eqs.~(\ref{2.11}) and
(\ref{2.17})--(\ref{2.21}), we find that the Fourier integral can be
expressed as 
\begin{eqnarray}
I(v,x_{\rm min},x_{\rm max}) &=& \frac{4\sqrt{30\pi}}{15}\, e^{-i\pi/4} 
v^{-7/2} \biggl[ 1 + \frac{92}{45}\, iv^5 
\nonumber \\ & & \mbox{}
+ O(v^{10}) + R(v,x_{\rm min},x_{\rm max}) \biggr],
\label{2.22}
\end{eqnarray}
where 
\begin{eqnarray}
R(v,x_{\rm min},x_{\rm max}) &=& -\frac{12}{\sqrt{30\pi}}\, e^{-i\pi/4} 
v^{7/2} \Bigl[ e^{-i\phi_{\rm min}} g(v, x_{\rm min}) 
\nonumber \\ & & \mbox{}
+ e^{-i\phi_{\rm max}} g(v, x_{\rm max}) \Bigr] .
\label{2.23}
\end{eqnarray}
Substituting this into Eq.~(\ref{2.7}), treating $R$ and the $O(v^5)$
term in Eq.~(\ref{2.22}) as small quantities, we arrive at the
following expression for the Fourier transform:
\begin{equation}
\tilde{h}(f) = \tilde{h}_{\rm spa}(f) (1 + \delta A_{\rm w})
e^{i(\delta \psi + \delta \psi_{\rm w})}.
\label{2.24}
\end{equation}
Here,
\begin{equation}
\tilde{h}_{\rm spa}(f) = \frac{\sqrt{30\pi}}{24}\, 
\frac{Q {\cal M}^2}{r}\, v^{-7/2} e^{i \psi},
\label{2.25}
\end{equation}
with
\begin{equation}
\psi(v) = 2\pi f t_c - \Phi_c - \frac{\pi}{4} + \frac{3}{128v^5}, 
\label{2.26}
\end{equation}
is the stationary-phase approximation to the Fourier transform. Our
calculation reveals the existence of two types of correction
terms. The first is
\begin{equation}
\delta \psi(v) = \frac{92}{45}\, v^5 + O(v^{10}),
\label{2.27}
\end{equation}
which represents a small, but steadily growing phase drift. Notice
that $\delta \psi$ is of order $O(v^{10})$ relative to $\psi$. The
other correction terms come as a consequence of the abrupt cutoffs
imposed at $F = F_{\rm min}$ and $F = F_{\rm max}$. They are
\begin{eqnarray}
\delta A_{\rm w}(v) &=& -\frac{12}{\sqrt{30\pi}}\, v^{7/2} \Bigl[ 
g(v, x_{\rm min}) \cos(\phi_{\rm min} + {\textstyle \frac{\pi}{4}})
\nonumber \\ & & \mbox{}
+ g(v, x_{\rm max}) \cos(\phi_{\rm max} + {\textstyle \frac{\pi}{4}})
\Bigr]
\label{2.28}
\end{eqnarray}
and
\begin{eqnarray}
\delta \psi_{\rm w}(v) &=& \frac{12}{\sqrt{30\pi}}\, v^{7/2} \Bigl[ 
g(v, x_{\rm min}) \sin(\phi_{\rm min} + {\textstyle \frac{\pi}{4}})
\nonumber \\ & & \mbox{}
+ g(v, x_{\rm max}) \sin(\phi_{\rm max} + {\textstyle \frac{\pi}{4}})
\Bigr].
\label{2.29}
\end{eqnarray}
Notice that $\delta A_{\rm w}$ represents an amplitude modulation,
while $\delta \phi_{\rm w}$ is a phase modulation; both oscillate as a
function of frequency. The suffix ``w'' indicates that these corrections 
are associated with ``windowing''.

\section{Comparison with discrete Fourier transform}

The preceding analysis reveals that apart from windowing issues, 
the stationary-phase approximation to the Fourier transform is 
extremely accurate: Apart from the modulations $\delta A_{\rm w}$
and $\delta \psi_{\rm w}$, $\tilde{h}_{\rm spa}(f)$ differs from
$\tilde{h}(f)$ only by a small phase drift $\delta \psi$ of relative  
order $v^{10}$. In this section, we firm up this conclusion by
comparing $\tilde{h}_{\rm spa}(f)$ to $\tilde{h}_{\rm fft}(f)$, the
discrete Fourier transform of the function $h(t)$.

The discrete Fourier transform is evaluated by Fast Fourier Transform
(FFT), using the routines of {\it Numerical Recipes} \cite{19}. The 
time series is prepared as follows. 

We begin with $h(t)$ as displayed in Eq.~(\ref{1.1}), with the
irrelevant factor $Q {\cal M}/r$ set to unity. Thus,
\begin{equation}
h(t) = (\pi {\cal M} F)^{2/3} e^{-i \Phi},
\label{3.1}
\end{equation}
where $F(t)$ and $\Phi(t)$ are given by Eqs.~(\ref{1.2}) and
(\ref{1.3}). This function is assumed to be nonzero only in the
interval $F_{\rm min} < F(t) < F_{\rm max}$. The duration of the
signal is 
\begin{equation}
T = t(F_{\rm max}) - t(F_{\rm min}),
\label{3.2}
\end{equation}
while the total number of wave cycles is 
\begin{equation}
{\cal N} = \frac{1}{2\pi}\bigl[ \Phi(F_{\rm max}) -
\Phi(F_{\rm min}) \bigr].
\label{3.3}
\end{equation}
The values of $h(t)$ at the endpoints $t(F_{\rm min})$ and $t(F_{\rm
  max})$ do not agree. This is a potential difficulty for the FFT,
which considers the signal to be periodic with period $T$. To remedy
this, we prepare our time series by padding $h(t)$ with zeros on both
sides. More precisely, we let $h(t)$ be zero in the interval $0 < t <
T$, be equal to the expression (\ref{3.1}) in the interval $T < t <
2T$, and be zero again in the interval $2T < t < 4T$. Thus, the
effective duration of the time series is four times the duration of
the actual signal. We choose the value of the parameters $t_c$ and
$\Phi_c$ such that $t(F_{\rm min}) \equiv T$ and $\Phi(F_{\rm min})
\equiv 0$.  This particular padding of the time series is somewhat
{\it ad hoc} (it has not been carefully chosen to be the smallest
needed to make negligible the circular correlations of the FFT), but
as we shall see the final results are very good and optimization would
be redundant.

The zero-padded function $h(t)$ is discretely sampled at times 
$t_k = k \Delta t$, where $k = 0,1,\ldots,4N-1$ and $\Delta t = 
4T/(4N - 1)$, with $4N$ denoting the number of sampled points. The
Nyquist frequency \cite{19} is given by $f_{\rm Ny} = 1/(2\Delta t)$, 
and $N$ must be adjusted so that $f_{\rm Ny} > F_{\rm max}$. The FFT 
returns the Fourier transform of the time series, discretely sampled 
in the frequency domain. The frequency resolution is $\Delta f = 
(4N \Delta t)^{-1}$.

We denote the discrete Fourier transform of the zero-padded time
series $h(t)$ by $\tilde{h}_{\rm fft}(f)$, and we wish to compare
this to $\tilde{h}_{\rm spa}(f)$, the stationary-phase approximation
given by Eqs.~(\ref{2.25}) and (\ref{2.26}). (It is understood that the
correct values for $t_c$ and $\Phi_c$ are substituted in these equations.)
To do this we define the relative amplitude $A_{\rm rel}$ and relative
phase $\psi_{\rm rel}$ by
\begin{equation}
A_{\rm rel}(f) = \mbox{mod} 
\Bigl[ \tilde{h}_{\rm fft}(f)/\tilde{h}_{\rm spa}(f) \Bigr]
\label{3.4}
\end{equation}
and 
\begin{equation}
\psi_{\rm rel}(f) = \mbox{arg} 
\Bigl[ \tilde{h}_{\rm fft}(f)/\tilde{h}_{\rm spa}(f) \Bigr],
\label{3.5}
\end{equation}
where $\mbox{mod}(z) = r$ is the modulus of the complex number
$z = r e^{i\theta}$, while $\mbox{arg}(z) = \theta$ is its argument. 
If $\tilde{h}_{\rm fft}(f)$ and $\tilde{h}_{\rm spa}(f)$ were in
perfect agreement, then $A_{\rm rel} = 1$ and $\psi_{\rm rel} = 0$.

Figure 1 shows plots of $\mbox{mod}(\tilde{h}_{\rm fft})$ and
$\mbox{mod}(\tilde{h}_{\rm spa})$ as functions of $f$ for a signal
prepared with ${\cal M} = 1.25\ M_\odot$ in the interval between
$F_{\rm min} = 40\ \mbox{Hz}$ and $F_{\rm max} = 60\ \mbox{Hz}$.
(Such a narrow band is not at all typical for inspiral signals sought
by any interferometer; we use it merely to exaggerate the errors
caused by the stationary phase approximation to the level where they
are visible.)  The duration of such a signal is $T = 15.8\ \mbox{s}$,
and the total number of wave cycles is ${\cal N} = 749$. The FFT was
taken with $N = 2048$, giving a Nyquist frequency of $f_{\rm Ny} = 65\ 
\mbox{Hz}$. The figure shows that the agreement is not perfect: While
the stationary-phase approximation seems to give the mean curve, the
discrete Fourier transform displays oscillations about the mean, and
these grow near the boundary points, $f = F_{\rm min}$ and $f = F_{\rm
  max}$. We shall argue that the discrepancy is entirely caused by
windowing.

\begin{figure}
\special{hscale=35 vscale=35 hoffset=-25.0 voffset=10.0
         angle=-90.0 psfile=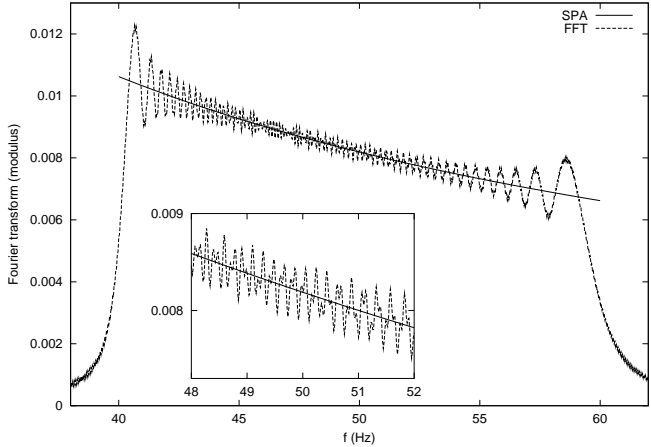}
\vspace*{2.6in}
\caption{The solid curve labeled ``SPA'' is a plot of 
$\mbox{mod}[\tilde{h}_{\rm spa}(f)]$, and the dashed curve labeled
``FFT'' is a plot of $\mbox{mod}[\tilde{h}_{\rm fft}(f)]$. The inset 
shows the same curves in a smaller frequency interval.}
\end{figure}

In Fig.~2 we show plots of $A_{\rm rel}(f)$ --- defined in
Eq.~(\ref{3.4}) --- and $1 + \delta A_{\rm w}(f)$ --- defined in 
Eq.~(\ref{1.7}). We recall that $\delta A_{\rm w}(f)$ represents the
amplitude modulation induced by windowing. The near-perfect agreement  
between $A_{\rm rel}(f)$ and $1 + \delta A_{\rm w}(f)$ shows that any
discrepancy between the discrete Fourier transform and the
stationary-phase approximation must be attributed to windowing. This
conclusion is confirmed by Fig.~3, which shows plots of 
$\psi_{\rm rel}(f)$ --- defined in Eq.~(\ref{3.5}) --- and $
\delta \psi_{\rm w}(f)$ --- defined in Eq.~(\ref{1.9}). Again we see a
near-perfect agreement, indicating that windowing accounts fully for
any discrepancy between $\tilde{h}_{\rm spa}(f)$ and 
$\tilde{h}_{\rm fft}(f)$.

\begin{figure}
\special{hscale=35 vscale=35 hoffset=-25.0 voffset=10.0
         angle=-90.0 psfile=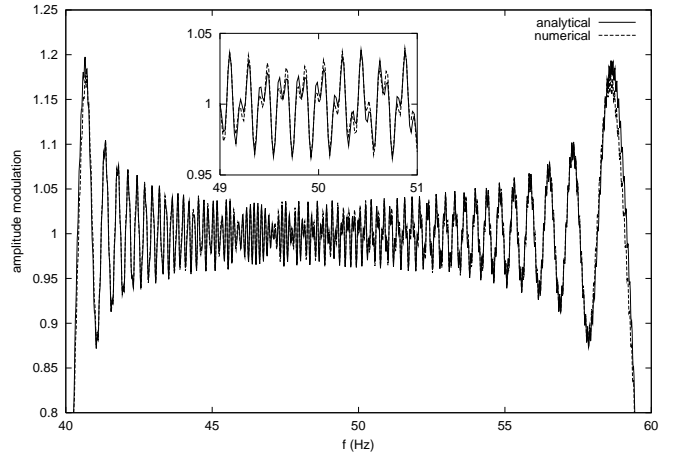}
\vspace*{2.6in}
\caption{The solid curve labeled ``analytical'' is a plot of 
$1 + \delta A_{\rm w}(f)$, and the dashed curve labeled ``numerical''
is a plot of $A_{\rm rel}(f)$. The inset shows the same curves in a 
smaller frequency interval.}
\end{figure}

\begin{figure}
\special{hscale=35 vscale=35 hoffset=-25.0 voffset=10.0
         angle=-90.0 psfile=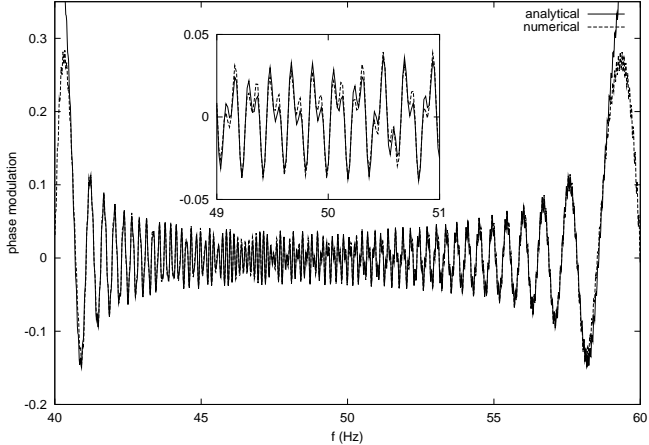}
\vspace*{2.6in}
\caption{The solid curve labeled ``analytical'' is a plot of 
$\delta \psi_{\rm w}(f)$, and the dashed curve labeled ``numerical''
is a plot of $\psi_{\rm rel}(f)$. The inset shows the same curves 
in a smaller frequency interval. Notice that the error in the phase
does not accumulate, and that it is always much smaller than $2\pi$.}
\end{figure}

Equations (\ref{1.7}) and (\ref{1.9}) give approximate expressions for
$\delta A_{\rm w}(f)$ and $\delta \psi_{\rm w}(f)$, and we should  
expect that in some situations, there could be noticeable differences
between these quantities and the numerically-determined 
$A_{\rm rel}(f)$ and $\psi_{\rm rel}(f)$. Figure 4 indicates that such
is indeed the case when the frequency interval is expanded. Here, the
signal is prepared with the same chirp mass as before, but the
frequency interval is now between $F_{\rm min} = 40\ \mbox{Hz}$ and 
$F_{\rm max} = 1300\ \mbox{Hz}$; other relevant quantities are listed
in Table 1 below. Although the agreement is no longer near-perfect, it
is still remarkably good, and this re-enforces our claim that any
discrepancy between the discrete Fourier transform and the
stationary-phase approximation is entirely an artifact of
windowing. We have verified that the intrinsic correction to the 
stationary-phase approximation, $\delta \psi(f)$ given by
Eq.~(\ref{1.6}), is irrelevant in this frequency interval: This 
phase shift is just too small to be noticeable at these frequencies. 

\begin{figure}
\special{hscale=35 vscale=35 hoffset=-25.0 voffset=10.0
         angle=-90.0 psfile=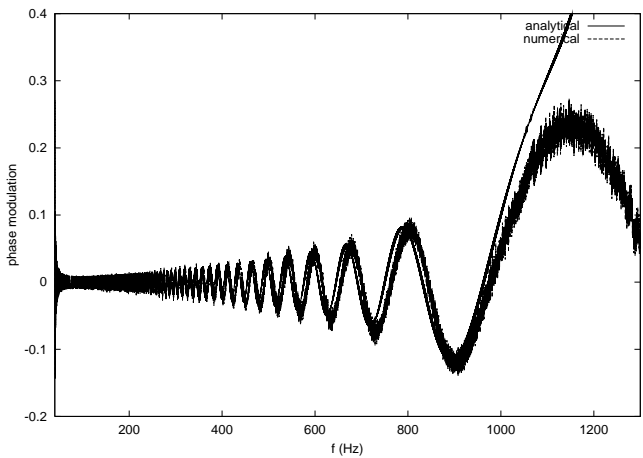}
\vspace*{2.6in}
\caption{The thin, solid curve labeled ``analytical'' is a plot of 
$\delta \psi_{\rm w}(f)$, and the thick, dashed curve labeled
``numerical'' is a plot of $\psi_{\rm rel}(f)$. Notice that here also,
the error in the phase does not accumulate, and is always much smaller
than $2\pi$.}
\end{figure}

\section{Overlap integral}

We have shown that any discrepancy between $\tilde{h}_{\rm fft}(f)$  
and $\tilde{h}_{\rm spa}(f)$ can be fully attributed to windowing,
which induces amplitude and phase modulations in the Fourier
transform. In this section we show that these modulations have no
significant effect on operations associated with matched filtering.  

The standard theory of matched filtering predicts that the loss
of signal-to-noise ratio incurred when filtering a signal 
$\tilde{h}_{\rm fft}(f)$ with a filter $\tilde{h}_{\rm spa}(f)$ 
is equal to \cite{8}
\begin{equation}
{\cal O} = \frac{ \bigl( h_{\rm fft} | h_{\rm spa} \bigr)}
{ \sqrt{ \bigl( h_{\rm fft} | h_{\rm fft} \bigr)
\bigl( h_{\rm spa} | h_{\rm spa} \bigr) } }.
\label{4.1}
\end{equation}
We will refer to this quantity as the {\it overlap} between the two
expressions for the signal's Fourier transform. An overlap close to
unity indicates that the filter is an accurate representation of the
signal, and that using this filter in analyzing the data will return
the largest possible signal-to-noise ratio. We use the notation
\begin{equation}
(a|b) = 2 \int_{F_{\rm min}}^{F_{\rm max}} 
\frac{\tilde{a}^*(f) \tilde{b}(f) + \tilde{a}(f) 
\tilde{b}^*(f)}{S_n(f)}\, df,
\label{4.2}
\end{equation}
where $S_n(f)$ is the spectral density of the detector noise. 

Notice that the point of view expressed here is that
$\tilde{h}_{\rm fft}(f)$ is an exact representation of the
signal's Fourier transform, while $\tilde{h}_{\rm spa}(f)$
is an approximate filter. However, because of its symmetry
in these quantities, $\cal O$ also represents the loss in
signal-to-noise ratio incurred when filtering a signal 
$\tilde{h}_{\rm spa}(f)$ with a filter $\tilde{h}_{\rm fft}(f)$. 
This is the opposite point view, in which the stationary-phase
approximation is viewed as an exact representation of the
Fourier transform.  

We evaluate the integrals in Eq.~(\ref{4.1}) by turning them
into discrete sums, using the sampled frequencies $f_k = k\Delta f$  
returned by the FFT. Thus, $\int \alpha(f)\, df \approx
\sum_k \alpha(f_k) \Delta f$. It is sufficient for our purposes to use
a simple analytic model for the noise's spectral density. We choose a
noise curve that roughly mimics the expected noise spectrum of the initial
LIGO detector, and set \cite{5}
\begin{equation}
S_n(f) = S_0 \Bigl[ (f_0/f)^4 + 2 + 2(f/f_0)^2 \Bigr]
\label{4.3}
\end{equation}
for $f > 40\ \mbox{Hz}$, with $f_0 = 200\ \mbox{Hz}$. The 
value of $S_0$ is irrelevant for our purposes, and $S_n(f)$ 
is taken to be infinite below 40 Hz. 

\begin{table}
\caption{The last column gives the overlap $\cal O$ between stationary
phase and FFT waveforms prepared with (identical)
chirp mass $\cal M$ given in solar masses in
the first column, initial frequency $F_{\rm min}$ given in Hz in the
second column, and final frequency $F_{\rm max}$ given in Hz in the
third column. The waveforms have a duration $T$ given in seconds in
the fourth column and a number of wave cycles $\cal N$ given in
the fifth column. The number of sampled times is $4N$, where $N$ is
given in the sixth column, corresponding to a Nyquist frequency 
$f_{\rm Ny}$ given in Hz in the seventh column. In all cases the
value of $F_{\rm max}$ is sufficiently large that the contribution
to the overlap from higher frequencies can be neglected (assuming the
initial LIGO noise spectrum given in the text).}
\begin{tabular}{dddddddd}
${\cal M}$ & $F_{\rm min}$ & $F_{\rm max}$ & 
$T$ & $\cal N$ & $N$ & $f_{\rm Ny}$ & $\cal O$ \\
$M_{\odot}$ & Hz & Hz & s & & & Hz & \\
\hline
      &    &      &      &      &          &      &        \\
1.00  & 40 &\ 900 & 34.6 & 2200 & $2^{16}$ &\ 948 & 0.9999 \\
1.25  & 40 & 1300 & 23.8 & 1521 & $2^{16}$ & 1375 & 0.9998 \\
1.50  & 40 & 1300 & 17.6 & 1122 & $2^{16}$ & 1863 & 0.9997 \\
1.75  & 40 & 1200 & 13.6 &\ 868 & $2^{15}$ & 1204 & 0.9997 \\
2.00  & 40 & 1500 & 10.9 &\ 695 & $2^{15}$ & 1505 & 0.9996 \\
2.25  & 40 &\ 900 & 8.9  &\ 570 & $2^{14}$ &\ 916 & 0.9994 \\
2.50  & 40 & 1000 & 7.5  &\ 478 & $2^{14}$ & 1091 & 0.9994 \\
2.75  & 40 & 1200 & 6.4  &\ 409 & $2^{14}$ & 1279 & 0.9994 \\
3.00  & 40 & 1400 & 5.5  &\ 354 & $2^{14}$ & 1479 & 0.9994 \\
10.00 & 40 & 1300 & 0.7 &\ \ 48 & $2^{11}$ & 1374 & 0.9972 
\end{tabular}
\end{table}

The overlap $\cal O$ is calculated for a number of chirp masses. The
results are displayed in Table 1. The conclusion is clear: For all
cases, ${\cal O} > 0.997$, indicating that the amplitude and phase
modulations have very little effect on matched filtering. In view of
the fact that the modulations oscillate and never get large,
especially away from $F_{\rm min}$ and $F_{\rm max}$ where the
instrument is most sensitive, this is the expected conclusion.

It should be noted that for most of the binaries listed in Table I,
the adopted value for $F_{\rm max}$ {\it exceeds} the frequency at
which the last stable orbit is expected to be found. (For equal-mass
systems, this frequency is given approximately by $1910 (M_\odot/{\cal
M})\ \mbox{Hz}$.) Our gravitational-wave signals are therefore not
realistic at high frequencies, something which is already made clear
by the fact that we do not incorporate post-Newtonian corrections into
our waveforms. Since our purpose here is simply to establish the
validity of the stationary-phase approximation, this lack of realism
is not too important. To produce a realistic waveform at high
frequencies, in a regime where the slow inspiral gives way to a rapid
merger of the two stars, is still an open and challenging problem in
gravitational-wave research. 

\section*{Acknowledgments}

The authors express gratitude toward P.R.~Brady, A.~Gopakumar,
B.S.~Sathyaprakash, and A.G.~Wiseman for discussions. The work in
Guelph was supported by the Natural Sciences and Engineering Research
Council of Canada. The work in Pasadena was supported by the National
Science Foundation under the grant PHY-9424337, and under the NSF
graduate program. E.P. wishes to thank Kip Thorne for his kind
hospitality at the California Institute of Technology, where part of
this work was carried out.

\end{document}